\documentclass[twocolumn,showpacs,amsmath,amssymb]{revtex4}
\usepackage{graphicx}
\begin{document}
\title{Modified equation of state, scalar field, and bulk viscosity in Friedmann universe}
\author{Jie Ren$^2$}
\altaffiliation{}\email{jrenphysics@hotmail.com} 
\author{Xin-He Meng$^{1,2}$}
\altaffiliation{}\email{xhm@nankai.edu.cn (Nankai -Univ.)}
\affiliation{$^{1}$CCAST (World Lab), P.O.Box 8730, Beijing 100080,
China\\$^{2}$Department of physics, Nankai University, Tianjin
300071, China (post address)}
\date{\today}
\begin{abstract}
A generalized dynamical equation for the scale factor of the
universe is proposed to describe the cosmological evolution, of
which the $\Lambda$CDM model is a special case. It also provides a
general example to show the equivalence of the modified equation
of state (EOS) and a scalar field model. In the mathematical
aspect, the EOS, the scalar field potential $V(\varphi)$, and the
scale factor $a(t)$ all have possessed analytical solutions. Such
features are due to a simple form invariance of the equation
inherited which determines the Hubble parameter. From the physical
point of view, this dynamical equation can be regarded as the
$\Lambda$CDM model with bulk viscosity, an existence content in
the universe. We employ the SNe Ia data with the parameter
$\mathcal{A}$ measured from the SDSS data and the shift parameter
$\mathcal{R}$ measured from WMAP data to constrain the parameters
in our model. The result is that the contribution of the bulk
viscosity, accumulated as an effective dark energy  responsible
for the current cosmic accelerating expansion, is made approximately
ten percent to that of the cosmological constant.
\end{abstract}
\maketitle

\section{Introduction}
The cosmological observations have provided increasingly
convincing evidence that our universe is undergoing a late-time
cosmic acceleration expansion \cite{rie98,bah99}. In order to
explain the acceleration expansion, physicists have introduced a new
fluid, which possesses a negative enough pressure, called dark
energy. According to the observational evidence, especially from
the Type Ia Supernovae \cite{rie04,jas06} and WMAP satellite
missions \cite{ben03}, we live in a favored spatially flat
universe consisting approximately of $30\%$ dark matter and $70\%$
dark energy. The simplest candidate for dark energy is the
cosmological constant, but it has got the (in-)famous and serious
fine-tuning problem, while the also elusive dark matter candidate might be
a lightest and neutral supersymmetry particle with only gravity interaction.
Recently, a great variety of models are
proposed to describe the universe, partly such as scalar fields,
exotic equation of state, modified gravity, inhomogeneous
cosmology model and super horizon backreaction\cite{sm}. It is also instructive
to investigate by concrete models
the possibility that the both mysterious dark components, dark matter and dark energy,
may be two facets of a single complicated fluid, that is, they originate from
a unified dark fluid in the universe.  From Einstein equation we know that the right
hand side is intimately related to the geometry or gravity, and on the other side
the energy-momentum tensor also involves a quantum field vacuum. According
to Ref.~\cite{cap05a}, it is tempting to investigate the
properties of cosmological models starting from the equation of
state (hereafter EOS) of dark energy directly and by testing
whether a given EOS is able to give rise to cosmological models
reproducing the available dataset. We hope the situation will be
improved with the new generation of more precise observational
data upcoming.

The observational constraints indicate that the current EOS
parameter $w=p/\rho$ is around $-1$ \cite{rie04}, quite probably
below $-1$, which is called the phantom region and even more
mysterious in the cosmological evolution processes. In the
standard model of cosmology, if the $w<-1$, the universe shows to
possess the future finite singularity called Big Rip
\cite{cal03,noj05a}. Several ideas are proposed to prevent the big
rip singularity, like by introducing quantum effects terms in the
action \cite{eli04,noj04b}. Based on the motivations of
time-dependent viscosity and modified gravity, the Hubble
parameter dependent EOS is considered in
Ref.~\cite{noj05b,cap05a}. The Hubble parameter dependent term in
this EOS can drive the phantom barrier being crossed in an easier
way \cite{bre05a,noj05b}. Different choices of the parameters may
lead to several fates to the cosmological evolution \cite{ren05}.
Recently, the equivalence between the modified EOS, the scalar
field model, and the modified gravity is demonstrated in
Ref.~\cite{cap05b}, with a general method to calculate the
potential of the scalar field for a given EOS presented.

In Refs.~\cite{zim96,col96,chi97}, the bulk viscosity in cosmology
has been studied in various aspects. Dissipative processes are
thought to be present in any realistic theory of the evolution of
the universe. In the early universe, the thermodynamics is far from
equilibrium, the viscosity should be concerned in the studies of the
cosmological evolution. Even in the later cosmic evolution stage,
for example, the temperature for the intergalactic medium (IGM), the
baryonic gas, generally is about $10^4$K to $10^6$K and the
complicated IGM is rather non-trivial. The sound speed $c_s$ in the
baryonic gas is only a few km/s to a few tens km/s and the Jeans
length $\lambda$ yields a term as an effective viscosity $c_s
\lambda$. On the other hand, the bulk velocity of the baryonic gas
is of the order of hundreds km/s \cite{pb}. So it is helpful to
consider the viscosity element in the later cosmic evolution. It is
well known that in the framework of Fredmann-Robertson-Walker (FRW)
metric, the shear viscosity \cite{jaf05} has no contribution in the
energy momentum tensor, and the bulk viscosity behaves like an
effective pressure. Moreover, the cosmic viscosity here can also be
regarded as an effective quantity as caused mainly by the
non-perfect cosmic contents interactions and may play a role as a
dark energy candidate \cite{zim01}.

In this letter, we show that in the framework of Friedmann
universe, the general EOS
\begin{displaymath}
p=(\gamma-1)\rho+p_0+w_{H}H+w_{H2}H^2+w_{dH}\dot{H},
\end{displaymath}
corresponds to a scalar field model whose potential has got the
form
\begin{displaymath}
V(\varphi)=V_0(e^{-\beta\varphi} +C_1e^{-\beta\varphi/2}+C_2).
\end{displaymath}
We will present analytically that the equation for the scale
factor derived from the above EOS or scalar field model is more
general than the $\Lambda$CDM model can show, and it has possessed
an exact solution. The natural interpretation of this model is
involved to the bulk viscosity. Concerning on the different forms
of the bulk viscosity coefficient, we propose three parameterized
$H(z)$ relations and use the observational data to constrain the
parameters.

This paper is organized as follows: In the next section we give the
generalized dynamical equation for the scale factor and show a
transformation to reduce the dynamical equation of the scale factor
$a(t)$ into a linear differential equation. In Sec. III we
demonstrate that the EOS is corresponding to a scalar field whose
potential can be exactly solved. In Sec. IV we find that there
exists a form invariance related to the variable cosmological
constant which satisfies a renormalization equation. In Sec. V we
use the SNe Ia data with the parameters $\mathcal{A}$ from large scale
survey and shift $\mathcal{R}$ from cosmic background radiation data
to constrain our model. Finally, we present our conclusions and discussions in the
last section.

\section{Viscous dark fluid described by an effective EOS}
We consider the FRW metric in the flat space geometry ($k$=0) as the
case favored by WMAP data
\begin{equation}
ds^2=-dt^2+a(t)^2(dr^2+r^2d\Omega^2),
\end{equation}
and assume that the cosmic fluid possesses a bulk viscosity $\zeta$.
The energy-momentum tensor can be written as
\begin{equation}
T_{\mu\nu}=\rho U_\mu U_\nu+(p-\zeta\theta)H_{\mu\nu},
\end{equation}
where in comoving coordinates $U^\mu=(1,0,0,0)$, and
$H_{\mu\nu}=g_{\mu\nu}+U_\mu U_\nu$ \cite{bre02}. By defining the
effective pressure as $\tilde{p}=p-\zeta\theta$ and from the
Einstein equation $R_{\mu\nu}-\frac{1}{2}g_{\mu\nu}R=8\pi
GT_{\mu\nu}$, we obtain the Friedmann equations
\begin{subequations}
\begin{eqnarray}
\frac{\dot{a}^2}{a^2} &=& \frac{8\pi G}{3}\rho\label{eq1},\\
\frac{\ddot{a}}{a} &=& -\frac{4\pi
G}{3}(\rho+3\tilde{p})\label{eq2}.
\end{eqnarray}
\end{subequations}
The conservation equation for energy, $T^{0\nu}_{;\nu}$, yields
\begin{equation}
\dot{\rho}+(\rho+\tilde{p})\theta=0,
\end{equation}
where $\theta=U^\mu_{;\mu}=3\dot{a}/a$.

In our previous work \cite{ren05}, we have considered the
following EOS form with the same notations
\begin{equation}
p=(\gamma-1)\rho+p_0+w_{H}H+w_{H2}H^2+w_{dH}\dot{H},\label{eq:eos}
\end{equation}
where $H$ is the Hubble parameter and $w_{x}$ are their
corresponding coefficients. We have  obtained the exact solution to
the scale factor and showed that
\begin{equation}
\zeta=\zeta_0+\zeta_1\frac{\dot{a}}{a}+\zeta_2\frac{\ddot{a}}{\dot{a}}
\end{equation}
is equivalent to the form by using the above EOS. By defining
\begin{eqnarray}
\tilde{\gamma} &=& \frac{\gamma+(\kappa^2/3)w_{H2}}{1+(\kappa^2/2)w_{dH}},\label{eq:gamma}\\
\frac{1}{T_1} &=& \frac{-(\kappa^2/2)
w_H}{1+(\kappa^2/2)w_{dH}},\\
\frac{1}{T_2^2} &=& \frac{-(\kappa^2/2)p_0}{1+(\kappa^2/2)w_{dH}},\\
\frac{1}{T^2} &=& \frac{1}{T_1^2}+\frac{6\tilde{\gamma}}{T_2^2},
\end{eqnarray}
the dynamical equation of the scale factor $a(t)$ can be written as
\begin{equation}
\frac{\ddot{a}}{a}=-\frac{3\tilde{\gamma}-2}{2}\frac{\dot{a}^2}{a^2}
+\frac{1}{T_1}\frac{\dot{a}}{a}+\frac{1}{T_2^2}.\label{eq:main}
\end{equation}
The analytical solution for $\tilde{\gamma}\neq 0$ is given out as
\begin{eqnarray}
a(t)=a_0\left\{\frac{1}{2}\left(1+\tilde{\gamma}\theta_0
T-\frac{T}{T_1}\right)\exp\left[\frac{t-t_0}{2}\left(\frac{1}{T}
+\frac{1}{T_1}\right)\right]\right.\nonumber\\
\left.+\frac{1}{2}\left(1-\tilde{\gamma}\theta_0
T+\frac{T}{T_1}\right)\exp\left[-\frac{t-t_0}{2}\left(\frac{1}{T}
-\frac{1}{T_1}\right)\right]\right\}^{2/3\tilde{\gamma}}.\label{eq:a}
\end{eqnarray}
For the case $\tilde{\gamma}=0$, the solution is
\begin{equation}
a(t)= a_0\exp\left[\left(\frac{1}{3}\theta_0T_1+\frac{T_1^2}{T_2^2}
\right)\left(e^{(t-t_0)/T_1}-1\right)-\frac{T_1(t-t_0)}{T_2^2}\right].\label{eq:g0a}
\end{equation}
The five parameters in Eq.~(\ref{eq:eos}) are condensed to
three free parameters in Eq.~(\ref{eq:main}) or its solution, later
the best fit analyses in Sec. V enable us to obtain the physical
evolution of the universe.

It is interesting that there exists a transformation
\begin{equation}
y=a^{3\tilde{\gamma}/2}\label{eq:trans}
\end{equation}
to reduce Eq.~(\ref{eq:main}) to a linear differential equation of
$y(t)$
\begin{equation}
\ddot{y}-\frac{1}{T_1}\dot{y}-\frac{3\tilde{\gamma}}{2T_2^2}y=0,\label{eq:lin}
\end{equation}
which can be solved easily. The equation of $a(t)$ can be written as
a more general form
\begin{equation}
\frac{\ddot{a}}{a}=-\frac{3\tilde{\gamma}-2}{2}\frac{\dot{a}^2}{a^2}
+\frac{1}{T_1}\frac{\dot{a}}{a}+\frac{1}{T_2^2}+\frac{C}{a^{3\tilde{\gamma}/2}},\label{eq:gen}
\end{equation}
and we can also use the transformation  of Eq.~(\ref{eq:trans}) to
obtain the linear differential equation of the function $y(t)$
\begin{equation}
\ddot{y}-\frac{1}{T_1}\dot{y}-\frac{3\tilde{\gamma}}{2T_2^2}y-\frac{3\tilde{\gamma}}{2}C=0,
\end{equation}
We can check directly that the following transformation
\begin{equation}
y=a^{3\tilde{\gamma}/2}+CT_2^2
\end{equation}
can reduce the nonlinear Eq.~(\ref{eq:gen}) to its corresponding
linear Eq.~(\ref{eq:lin}). If $\tilde{\gamma}=4/3$, i.e., $w=1/3$,
Eq.~(\ref{eq:gen}) becomes
\begin{equation}
\frac{\ddot{a}}{a}=-\frac{\dot{a}^2}{a^2}
+\frac{1}{T_1}\frac{\dot{a}}{a}+\frac{1}{T_2^2}+\frac{C}{a^2},
\end{equation}
which can be interpreted as the radiation dominated universe when
the curvature of the universe is concerned.

\section{Scalar Field and Modified Gravity}
Starting from the action for the gravitational with the matter
fields, we can show that an EOS for the universe contents
corresponds to a scalar field model. Generally, for a given EOS,
the potential $V(\varphi)$ often has got no analytical solution.
Here we demonstrate that the correspondent scalar field model for
the EOS of Eq.~(\ref{eq:eos}) luckily  has possessed an exact
solution. First, we revisit the general procedure proposed in
Ref.~\cite{cap05b} to relate a scalar field model for a given EOS.

Starting from the action of the scalar-tensor theory
\begin{equation}
S=\int d^4 x\sqrt{-g}\left(\frac{1}{2\kappa^2}R
-\frac{1}{2}\omega(\phi)\partial_\mu\phi\partial^\mu\phi-V(\phi)\right),
\end{equation}
the energy density and the pressure are
\begin{equation}
\rho=\frac{1}{2}\omega(\phi)\dot{\phi}^2+V(\phi),~~
p=\frac{1}{2}\omega(\phi)\dot{\phi}^2-V(\phi).
\end{equation}
Combining the above equations and the Friedmann equations, one
obtains
\begin{equation}
\omega(\phi)\dot{\phi}^2=-\frac{2}{\kappa^2}\dot{H},~~V(\phi)=\frac{1}{\kappa^2}(3H^2+\dot{H}).
\end{equation}
The interesting case is that $\omega(\phi)$ and $V(\phi)$ are
determined in terms of a single function $f(\phi)$ as
\begin{equation}
\omega(\phi)=-\frac{2}{\kappa^2}f'(\phi),~~V(\phi)=\frac{1}{\kappa^2}(3f(\phi)^2+f'(\phi)).\label{eq:v}
\end{equation}
One can check that the special solution $\phi=t$, $H=f(t)$
satisfies the scalar-field equation. The following relations is
obtained in order to solve the function $f(\phi)$ for a given EOS,
\begin{equation}
\rho=\frac{3}{\kappa^2}f(\phi)^2,~~p=
-\frac{3}{\kappa^2}f(\phi)^2-\frac{2}{\kappa^2}f'(\phi)\label{eq:phi}.
\end{equation}
By defining a new field $\varphi$ as
\begin{equation}
\varphi=\int d\phi\sqrt{|\omega(\phi)|},\label{eq:varphi}
\end{equation}
the action can be rewritten as
\begin{equation}
S=\int d^4 x\sqrt{-g}\left(\frac{1}{2\kappa^2}R
\mp\frac{1}{2}\partial_\mu\varphi\partial^\mu\varphi-\tilde{V}(\varphi)\right).
\end{equation}
The energy density and the pressure is now given by
\begin{equation}
\rho=\pm\frac{1}{2}\dot{\varphi}^2+\tilde{V}(\varphi),~~
p=\pm\frac{1}{2}\dot{\varphi}^2-\tilde{V}(\varphi).
\end{equation}

Now we summarize the procedure. For a given EOS, by using
Eqs.~(\ref{eq:phi}), one can obtain an equation for $f(\phi)$;
then solving the equation gives $f(\phi)$. Using
Eqs.~(\ref{eq:v}), one obtains $\omega(\phi)$ and $V(\phi)$. And
by employing Eq.~(\ref{eq:varphi}) to transform the variable
$\phi$ to $\varphi$, finally, one obtains the
$\tilde{V}(\varphi)$. In Ref.~\cite{cap05b}, several examples are
presented such as $p=w\rho$, and we show in this section that a
more general form of Eq.~(\ref{eq:eos}) can also have had an
analytical solution of the potential.

\begin{figure}[]
\includegraphics{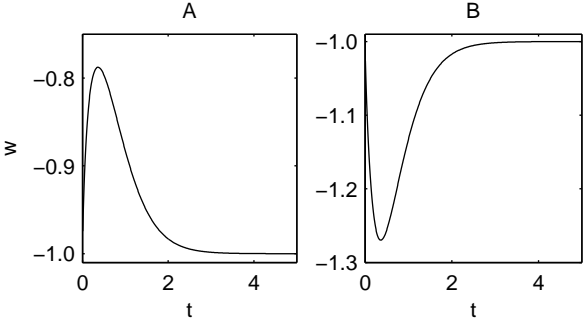}
\caption{\label{fig1} The $w$-$t$ relation of the quintessence case
(panel A) and the phantom case (panel B).}
\end{figure}

By using the Friedmann equations, Eq.~(\ref{eq:eos}) can be
rewritten as
\begin{equation}
p=(\tilde{\gamma}-1)\rho-\frac{2}{\sqrt{3}\kappa
T_1}\sqrt{\rho}-\frac{2}{\kappa^2 T_2^2}.\label{eq:vareos}
\end{equation}
The corresponding equation for $f(\phi)$ is
\begin{equation}
f'(\phi)=\frac{3\tilde{\gamma}}{2}f(\phi)^2-\frac{1}{T_1}f(\phi)-\frac{1}{T_2}.
\end{equation}
The solution of this equation is
\begin{equation}
f(\phi)=\alpha\coth(\frac{3\tilde{\gamma}}{2}\alpha\phi)+\frac{1}{3\tilde{\gamma}T_1},\label{eq:fphi}
\end{equation}
where
\begin{equation}
\alpha=-\sqrt{\frac{1}{9\tilde{\gamma}^2T_1^2}+\frac{2}{3\tilde{\gamma}T_2^2}}.
\end{equation}
Then we obtain the $\omega(\phi)$ and $V(\phi)$. The integration of
Eq.~(\ref{eq:varphi}) gives the relation between $\phi$ and
$\varphi$ as
\begin{equation}
\varphi=\frac{2}{\kappa}\sqrt{3|\tilde{\gamma}|}\ln\left|
\frac{\tanh(3\tilde{\gamma}\alpha\phi/2)}{\tanh(3\tilde{\gamma}\alpha\phi_0/2)}\right|,
\end{equation}
which gives
\begin{equation}
\coth(\frac{3\tilde{\gamma}}{2}\alpha\phi)=\coth(\frac{3\tilde{\gamma}}{2}\alpha\phi_0)
\exp(-\frac{\kappa}{2}\sqrt{3|\tilde{\gamma}|}\varphi).
\end{equation}
So substituting Eq.~(\ref{eq:fphi}) to Eq.~(\ref{eq:v}) and using
the above equation to transform the variable $\phi$ to $\varphi$, we
obtain
\begin{eqnarray}
\tilde{V}(\varphi) &=&
\frac{\alpha^2}{\kappa^2}\coth^2\left(\frac{3\tilde{\gamma}}{2}\alpha\phi_0\right)
\left[\frac{3(2-\tilde{\gamma})}{2}\exp\left(-\kappa\sqrt{3|\tilde{\gamma}|}\varphi\right)\right.\nonumber\\
&&
\left.-\frac{2}{\tilde{\gamma}T_1\alpha}\exp\left(-\frac{\kappa}{2}\sqrt{3|\tilde{\gamma}|}\varphi\right)
+\frac{1}{3\tilde{\gamma}^2T_1^2\alpha^2}+\frac{3\tilde{\gamma}}{2}\right].\label{potential}
\end{eqnarray}
This potential has got the form
\begin{equation}
\tilde{V}(\varphi)=V_0\left(e^{-\beta\varphi}+C_1e^{-\beta\varphi/2}+C_2\right).
\end{equation}
As a special case, $p=(\gamma-1)\rho$ is given out in
Ref.~\cite{cap05b}. Taking the limit of $T_1\to\infty$ and
$\alpha\to 0$ of Eq.~(\ref{potential}), and omitting the constant
term, we obtain
\begin{equation}
\tilde{V}(\varphi)=V_0e^{-\kappa\sqrt{|\gamma|}\varphi},
\end{equation}
where
\begin{equation}
V_0=\frac{2(2-\gamma)}{3\gamma^2\kappa^2\phi_0^2}.
\end{equation}
To see the dynamical effects of the scalar field, by
defining the EOS parameter
$w=(\pm\dot{\varphi}^2-\tilde{V})/(\pm\dot{\varphi}^2+\tilde{V})$,
the evolution of $w$ with respect to the time $t$ is illustrated in
Fig.~\ref{fig1}.

The action for a modified gravity in Einstein frame
\begin{equation}
S=\int d^4
x\sqrt{-g}\left(\frac{1}{2\kappa^2}R+\mathcal{L}_{\textrm{matter}}+f(R)\right)
\end{equation}
is shown to be related to a modified EOS \cite{cap05b}. In
Ref.~\cite{noj05b}, the following two equations are derived in the
framework of modified gravity,
\begin{equation}
0=-\frac{3}{\kappa^2}H^2+\rho-f(R)+6\left(\dot{H}+H^2-H\frac{d}{dt}\right)f'(R),\label{eq:rr}
\end{equation}
\begin{eqnarray}
0 &=& \frac{1}{\kappa^2}(2\dot{H}+3H^2)+p+f(R)\nonumber\\
&&
+2\left(-\dot{H}-3H^2+\frac{d^2}{dt^2}+2H\frac{d}{dt}\right)f'(R),\label{eq:pp}
\end{eqnarray}
where $R=6\dot{H}+12H^2$. If we adopt the approximation
$\dot{H}<<H^2$ \cite{noj05b}, and assume $f(R)=f_0\sqrt{R}$, by
combining Eqs.~(\ref{eq:rr}) and (\ref{eq:pp}), we can obtain
\begin{equation}
p=(\gamma-1)\rho-\frac{3\sqrt{3}}{2}f_0\gamma
H-\frac{3}{\kappa^2}\gamma H^2,
\end{equation}
which is a special case of Eq.~(\ref{eq:eos}). 
Additionally,
the scalar-tensor theory in Einstein frame is mathematically
equivalent to the modified gravity in Jordan frame \cite{cap05b}.
The conformal transformation $g_{\mu\nu}\to
e^{\pm\kappa\varphi\sqrt{\frac{2}{3}}}$ makes the kinetic term in
the action of scalar-tensor theory vanish, then one obtains the
Jordan frame action
\begin{equation}
S=\int d^4x\sqrt{-g}\left(\frac{e^{\pm
\kappa\varphi\sqrt{\frac{2}{3}}}}{2\kappa^2}R-e^{\pm
2\kappa\varphi\sqrt{\frac{3}{2}}}\tilde{V}(\varphi)\right).
\end{equation}
If the equation of motion of $\varphi$ can be solved as
$\varphi=\varphi(R)$, one obtains the action of the modified gravity
in Jordan frame. Thus, the Hilbert-Einstein action with an
additional term $f(R)$ has made effective contributions similar to those
caused by a scalar field, as well as a modified EOS.

\section{Variable Cosmological constant and Renormalization Group Equation}
The model of variable cosmological constant is another alternative
to explain the cosmic evolution, in order to overcome the serious
fine-tuning problem. In the framework of variable cosmological
constant model, the evolution of the scale factor is determined by
both the Friedmann equations and the renormalization group
equation (RGE) for the cosmological constant \cite{sha00,sha04},
\begin{eqnarray}
\frac{\ddot{a}}{a} &=&
-\frac{3\tilde{\gamma}-2}{2}\frac{\dot{a}^2}{a}
+\frac{3\tilde{\gamma}+1}{6}\Lambda,\label{eq:lambda}\\
\frac{d\Lambda}{d\ln\mu} &=& \frac{1}{4\pi^2}\sigma\mu^2M^2+....
\end{eqnarray}
Here we have already used the EOS $p=(\gamma-1)\rho$ to eliminate
the $\rho$ and $p$ in Friedmann equations. With the choice of the
renormalization scale $\mu=H$ \cite{sha00}, the variable $\Lambda$
is determined by
\begin{equation}
\frac{d\Lambda}{d\ln H}=\frac{1}{4\pi^2}\sigma H^2M^2.
\end{equation}
The solution is
\begin{equation}
\Lambda(t)=\Lambda_0+\xi[H(t)^2-H(t_0)^2]M_P^2,
\end{equation}
which has got the form of $\Lambda=C_0+C_2H^2$. Compared
Eq.~(\ref{eq:lambda}) with Eq.~(\ref{eq:main}), we find that if
Eq.~(\ref{eq:lambda}) is formly invariant under the transformation
\begin{equation}
\mu\to\mu+\delta\mu,
\end{equation}
the solution of RGE becomes
\begin{equation}
\Lambda=C_0+C_1H+C_2H^2.
\end{equation}
Substitute this result to Eq.~(\ref{eq:lambda}), we obtain a
equation which has the form of Eq.~(\ref{eq:main}). Especially, if
$\delta\mu=0$, then $C_1=0$. It is very interesting that
concerning on the bulk viscosity, modified EOS, scalar field model
, modified gravity, and the variable cosmological constant can be
described in one generally dynamical equation which determines the
scale factor.

Actually, the evolution equation of the Hubble parameter,
\begin{equation}
\dot{H}=-\frac{3\tilde{\gamma}}{2}H^2+\frac{1}{T_1}H+\frac{1}{T_2^2},
\end{equation}
has possessed a form invariance for $H\to H+H_0$, i.e.
\begin{equation}
\dot{H}=-\frac{3\tilde{\gamma}}{2}H^2+\left(\frac{1}{T_1}-2H_0\right)H
+\frac{1}{T_2^2}+H_0^2.
\end{equation}
It is this form invariance that gives several interesting features
of the model, such as both the $a(t)$ and $V(\varphi)$ have
analytical solutions for the general EOS of Eq.~(\ref{eq:eos}).

\section{A new content of the universe and data fitting}
In Eq.~(\ref{eq:main}), $1/T_2^2$ plays the role of the effective
cosmological constant. If $T_2\to\infty$, the $H-z$ relation is
\begin{equation}
H(z)=H_0[\Omega_m(1+z)^{3/2}+(1-\Omega_m)].
\end{equation}
We proposed a parameterized $H$-$z$ relation as the following form
\begin{equation}
H^2=H_0^2[\Omega_m(1+z)^{3\tilde{\gamma}}+\Omega_v(1+z)^{3/2}+1-\Omega_m
-\Omega_v],
\end{equation}
where $\Omega_v=2/(3\tilde{\gamma}T_1H_0)$ and
$1-\Omega_m-\Omega_v=2/(3\tilde{\gamma}T_2^2H_0^2)$. Note that
Eq.~(\ref{eq:main}) can have several interpretations \cite{ren05},
such as the "inflessence" model \cite{car05} when the first term and
third term alternatively dominate. The three terms in the right hand
side of Eq.~(\ref{eq:main}) are proportional to $H^2$, $H^1$, and
$H^0$, respectively. In the early times, the first term is dominant,
which may lead to inflation if $\tilde{\gamma}\sim 0$. In the medium
times, the second term dominates, which leads to deceleration if
$T_1<0$. In the late times as current stage, the third term is
dominant, which leads to cosmic acceleration expansion behaving as
the de Sitter universe if $T_2^2>0$. So, a single equation may
describe three epoches of the cosmological evolution. In this paper,
however, we only focus on the interpretation of bulk viscosity for
this model, as the viscous universe has been discussed for various
cosmology evolution stages with very naturally physical motivations.
And the viscosity and the dissipative processes in describing
physical universe have been studied in various aspects, for example
\cite{pri00,bre04,noj04c,cat05,bre05b}.

So far, the universe contents and their dynamical contributions are
listed in Table~\ref{tab:t1}.
\begin{table}
\caption{\label{tab:t1} }
\begin{ruledtabular}
\begin{tabular}{ccc}
$w$ & universe content & contribution to $H(z)^2$\\
\hline $-1$ & vacuum & $(1+z)^0$\\
$-1/2$ & (effective) viscosity & $(1+z)^{3/2}$\\
$-1/3$ & curvature & $(1+z)^2$\\
$0$ & dust & $(1+z)^3$\\
$1/3$ & radiation & $(1+z)^4$\\
$1$ & stiff matter & $(1+z)^6$
\end{tabular}
\end{ruledtabular}
\end{table}
Compared with the $\Lambda$CDM model, we use three cosmological
scenarios as parameterizations of the $H$-$z$ relations, which are
listed in Table~\ref{tab:t3}.
\begin{itemize}
\item (i) The $\Lambda$CDM model, the simplest model to explain
the dark energy. \item (ii) The viscosity model without
cosmological constant. \item (iii) The bulk viscosity is constant,
so that the bulk viscosity has got the dynamical effects of
$(1+z)^{3/2}$. \item (iv) The bulk viscosity has the form of
$\zeta=\zeta_0+\zeta_1\dot{a}/a$, where the constant term has the
dynamical effects of $(1+z)^{3/2}$, and the term proportional to
$H$ has an effect to change $\gamma$ to $\tilde{\gamma}$. (See
Eq.~(\ref{eq:gamma}) and Ref.~\cite{bre05a})
\end{itemize}

The observations of the SNe Ia have provided the direct evidence
for the cosmic accelerating expansion for our current universe.
Any model attempting to explain the acceleration mechanism should
be consistent with the SNe Ia data implying results, as a basic
requirement. The method of the data fitting is illustrated in
Refs.~\cite{ben05,gon05}. The observations of supernovae measure
essentially the apparent magnitude $m$, which is related to the
luminosity distance $d_L$ by
\begin{equation}
m(z)=\mathcal{M}+5\log_{10} D_L(z),
\end{equation}
where $D_L(z)\equiv(H_0/c)d_L(z)$ is the dimensionless luminosity
distance and
\begin{equation}
d_L(z)=\frac{c(1+z)}{H_0}\int_0^z\frac{1}{E(z')}dz'.
\end{equation}
Also,
\begin{equation}
\mathcal{M}=M+5\log_{10}\left(\frac{c/H_0}{1{\rm{Mpc}}}\right)+25,
\end{equation}
where $M$ is the absolute magnitude which is believed to be
constant for all supernovae of type Ia. We use the 157 golden
sample of supernovae Ia data compiled by Riess \textit{et al.}
\cite{rie04} to fit our model. The data points in these samples
are given in terms of the distance modulus
\begin{equation}
\mu_{obs}(z)\equiv m(z)-M_{obs}(z).
\end{equation}
The $\chi^2$ is calculated from
\begin{eqnarray}
\chi^2=\sum_{i=1}^{n}\left[\frac{\mu_{obs}(z_i)-\mathcal{M'}
-5\log_{10}D_{Lth}(z_i;c_\alpha)}{\sigma_{obs}(z_i)}\right]^2\nonumber\\
+\left(\frac{\mathcal{A}-0.469}{0.017}\right)^2+\left(\frac{\mathcal{R}-1.716}{0.062}\right)^2.
\end{eqnarray}
where $\mathcal{M'}=\mathcal{M}-M_{obs}$ is a free parameter and
$D_{Lth}(z_i;c_\alpha)$ is the theoretical prediction for the
dimensionless luminosity distance of a SNe Ia at a particular
distance, for a given model with parameters $c_\alpha$. The
parameter $\mathcal{A}$ is defined as \cite{eis05}
\begin{equation}
\mathcal{A}=\frac{\sqrt{\Omega_m}}{0.35}\left[\frac{0.35}{E(0.35)}\int_0^{0.35}
\frac{dz}{E(z)}\right]^{1/3},
\end{equation}
and the shift parameter $\mathcal{R}$ is \cite{wan04a,wan04b}
\begin{equation}
\mathcal{R}=\sqrt{\Omega_m}\int_0^{z_{ls}}\frac{dz}{E(z)}.
\end{equation}

\begin{figure}[]
\includegraphics{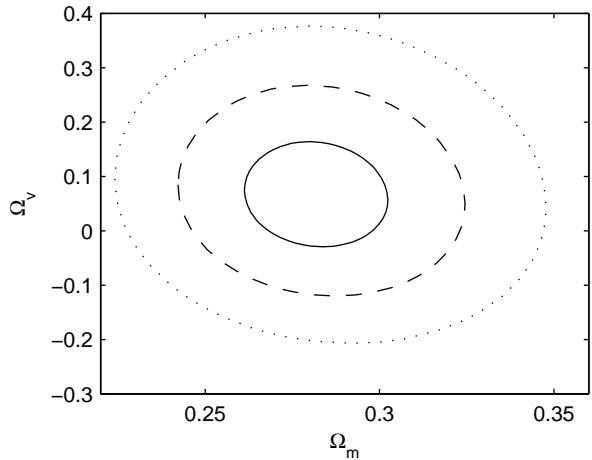}
\caption{\label{fig2} The $1\sigma$ (solid line), $2\sigma$ (dashed
line), and $3\sigma$ (dotted line) contour plots of model (iii).}
\end{figure}

\begin{figure}[]
\includegraphics{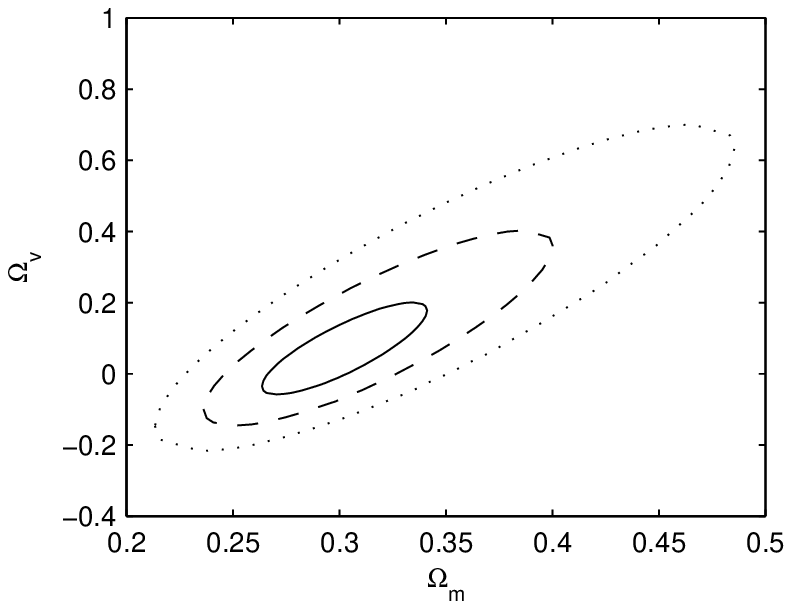}
\caption{\label{fig3} The $1\sigma$ (solid line), $2\sigma$ (dashed
line), and $3\sigma$ (dotted line) contour plots of
$\Omega_m$-$\Omega_v$ relation in model (iv).}
\end{figure}

\begin{figure}[]
\includegraphics{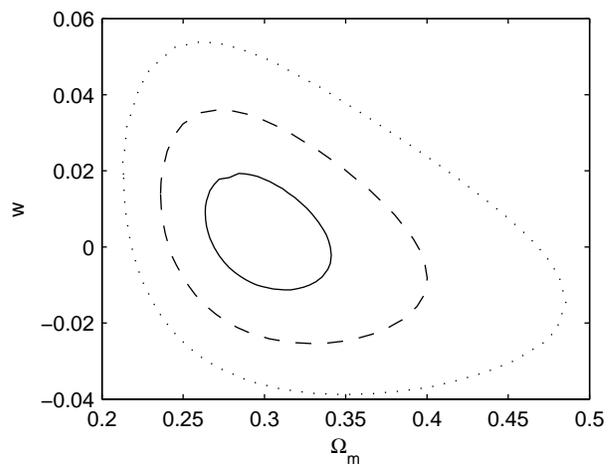}
\caption{\label{fig4} The $1\sigma$ (solid line), $2\sigma$ (dashed
line), and $3\sigma$ (dotted line) contour plots of $\Omega_m$-$w$
relation in model (iv).}
\end{figure}

We will consider the $\Lambda$CDM model for comparison and perform
a best-fit analysis with the minimization of the $\chi^2$, with
respect to $\mathcal{M'}$, $\Omega_m$, $\Omega_v$, and
$\tilde{\gamma}$. The results are summarized in
Table~\ref{tab:t2}.
\begin{table*}
\caption{\label{tab:t2} Summary of the best fit parameters}
\begin{ruledtabular}
\begin{tabular}{cccc}
& Models & best fit of parameters & $\chi^2$\\
\hline (i) & $H^2=H_0^2[\Omega_m(1+z)^3+1-\Omega_m]$
& $\Omega_m=0.283$ & 177.84\\
(ii) & $H=H_0^2[\Omega_m(1+z)^{3/2}+1-\Omega_m]$
& $\Omega_m=0.435$ & 319.78\\
(iii) &
$H^2=H_0^2[\Omega_m(1+z)^3+\Omega_v(1+z)^{3/2}+1-\Omega_m-\Omega_v]$
& ($\Omega_m,\Omega_v)=(0.281,0.065)$ & 177.38\\
(iv) &
$H^2=H_0^2[\Omega_m(1+z)^{3\tilde{\gamma}}+\Omega_v(1+z)^{3/2}+1-\Omega_m-\Omega_v]$
& ($\Omega_m,\Omega_v,\tilde{\gamma})=(0.298,0.053,1.004)$ & 177.32
\end{tabular}
\end{ruledtabular}
\end{table*}
From the results, we see that the bulk viscosity part has made
approximately $10\%$ contributions to that of the cosmological
constant. In model (ii), the cosmic acceleration expansion is due to
the bulk viscosity, without the cosmological constant, and with
comparing this model is rather disfavored. For model (iii),
Fig.~\ref{fig2} plots the likelihood contour of the parameters
$\Omega_m$ and $\Omega_v$. We can see that the bulk viscosity
contributes approximately $10\%$ of the cosmological constant. For
model (iv), Fig.~\ref{fig3} and Fig.~\ref{fig4} plot the likelihood
contours of $\Omega_m-\Omega_v$ and $\Omega_m-w$
($w=\tilde{\gamma}-1$), respectively.

\section{Conclusion and Discussion}
In conclusion, we have shown that several different approaches to
explain the current accelerating universe expansion can give the
same form of a dynamical equation for $a(t)$. Also in the sense that
different terms in the right hand side of the dynamical equation may
dominate correspondingly different periods we call the media
described by the general EOS a Dark Fluid. The case of the
$\Lambda$CDM model can be regarded as a special case of this
equation descriptions. In the framework of Friedmann universe, the
correspondences of the bulk viscosity, EOS, scalar field, modified
gravity, and variable cosmological constant are summarized in
Table~\ref{tab:t3}. 
The scalar field model as a prototype is intensely used to
study the dynamical behaviors for the scalar factor in the
literature because of its simplicity in formulation and treatment;
here we have provided a general and exact solution of
the scalar field which can be either quintessence or phantom.
Mathematically, this equation enjoys several interesting features,
such as the EOS, the potential $V(\varphi)$ of the scalar field
model, and the scale factor $a(t)$ all have got analytical
solutions. This equation can be transformed to a linear equation and
has possessed a form invariance related to the Hubble parameter.
Physically, the bulk viscosity is a most natural interpretation of
this model. We have proposed three cosmological scenarios which
contain the contributions from the bulk viscosity and used the 157
golden sample of SNe Ia data with the parameter $\mathcal{A}$ and
$\mathcal{R}$ to constrain our model. The special model that only
the bulk viscosity contributes the cosmic accelerating expansion
without the cosmological constant may be excluded. And  the
(effective) bulk viscosity has made approximately $10\%$
contributions to that of the cosmological constant.

The puzzling cosmic dark components: dark matter and dark energy, responsible mainly
for large scale structure formation of universe
and cosmic accelerating expansion as well as our universe evolution fate
as we now understand in the standard hot big bang and inflation models have challenged
our previous intelligence on the physical world. A unification picture
description for the two
elusive dark composites either from complicated fluid dynamics, modified gravity,
inhomogeneous cosmology or quantum
field models with introducing more degrees of freedom and supported by more
precious experiments like LHC and PLANCK in 2007 is certainly valuable for us to uncover the mysterious dark side of
the Universe, which even will bring us new knowledge on fundamental physics.

\begin{table}[h]
\caption{\label{tab:t3} The mathematical equivalence of different
approaches in cosmology ($C_i$ only denote the coefficients)}
\begin{ruledtabular}
\begin{tabular}{cc}
Bulk viscosity &
$\zeta=\zeta_0+\zeta_1\dot{a}/a+\zeta_2\ddot{a}/\dot{a}$\\
EOS & $p=(\gamma-1)\rho+p_0+w_{H}H+w_{H2}H^2+w_{dH}\dot{H}$\\
& (or $p=C_1\rho+C_2\sqrt{\rho}+C_3$)\\
Scalar field & $V(\varphi)=V_0(e^{-\beta\varphi} +C_1e^{-\beta\varphi/2}+C_2)$\\
Modified Gravity & $f(R)=f_0\sqrt{R}$ (by approximation)\\
Variable CC & $\Lambda=C_0+C_1H+C_2H^2$
\end{tabular}
\end{ruledtabular}
\end{table}

Note: In this brief presentation we find by numerical calculations
that parameters $\mathcal{A}$ and $\mathcal{R}$ respectively from
large scale survey and cosmic background radiation detections
significantly affect the fitting results for some models. For
example, if we do not use $\mathcal{A}$ and $\mathcal{R}$, the
minimum $\chi^2$ of Model (ii) is $179.36$, which is acceptable.
Furthermore, the best-fit results of Model (iii) are listed in
Table~\ref{tab:t4}. From this table, it is obvious that the result
by using the 157 golden samples (hereafter 157 as in the below
table) with parameters $\mathcal{A}$ and $\mathcal{R}$ is very
different from that by only using the 157 golden samples, for
which we are trying to figure out the physical reasons in late study.
It is very likely that we can obtain more accuracy results with high level
certainties by next year (2007)
PLANCK mission, continuous large scale structure and later large sample of SNe Ia survey, to which we are very
confident.
\begin{table}[h]
\caption{\label{tab:t4} Fitting results for the model
$H^2=H_0^2[\Omega_m(1+z)^3+\Omega_v(1+z)^{3/2}+1-\Omega_m-\Omega_v]$
with different data}
\begin{ruledtabular}
\begin{tabular}{cccc}
& $\Omega_m$ & $\Omega_v$ & $\chi^2$\\
\hline
$157$ & $0.330$ & $-0.583$ & $174.53$\\
$157+\mathcal{A}+\mathcal{R}$ & $0.281$ & $0.065$ & $177.38$\\
$157+\mathcal{A}$ & $0.300$ & $0.063$ & $177.36$\\
$157+\mathcal{R}$ & $0.303$ & $0.045$ & $177.34$\\
$\mathcal{A}+\mathcal{R}$ & $0.316$ & $0.117$ & $1.88\times 10^{-7}$\\
$\mathcal{A}$ & $0.283$ & $0.031$ & $1.34\times 10^{-9}$\\
$\mathcal{R}$ & $0.266$ & $0.033$ & $3.47\times 10^{-8}$
\end{tabular}
\end{ruledtabular}
\end{table}

\section*{ACKNOWLEDGEMENTS}
We thank Prof. S.D. Odintsov for the helpful comments with reading
the manuscript, and  Profs. I. Brevik and L. Ryder for lots of
discussions. This work is supported partly by NSF and Doctoral
Foundation of China.

\end{document}